\documentclass[journal]{IEEEtran}
\ifCLASSINFOpdf
\else
\fi

\usepackage{graphicx}
\usepackage{algorithm}
\usepackage{algpseudocode}

\usepackage{graphicx}
\usepackage[table]{xcolor}
\usepackage{multirow}
\usepackage{colortbl}
\usepackage{hhline}
\usepackage[para]{threeparttable}
\usepackage{cleveref}
\usepackage{tikz,colortbl}
\usepackage{makecell}
\usepackage{pdfpages}
\usepackage{tabu}

\begin{document}
%
\title{Leveraging AES Padding: dBs for Nothing and FEC for Free in IoT Systems}

\author{Jongchan Woo,~\IEEEmembership{Graduate Student Member,~IEEE}, Vipindev Adat Vasudevan,~\IEEEmembership{Member,~IEEE}, Benjamin D. Kim,~\IEEEmembership{Student Member,~IEEE}, Rafael G. L. D'Oliveira,~\IEEEmembership{Member,~IEEE}, Alejandro Cohen,~\IEEEmembership{Member,~IEEE}, Thomas Stahlbuhk, Ken R. Duffy,~\IEEEmembership{Senior Member,~IEEE} and Muriel Médard,~\IEEEmembership{Fellow,~IEEE}

\thanks{Jongchan Woo, Vipindev Adat Vasudevan, and Muriel Médard are with the Department of Electrical Engineering and Computer Science, Massachusetts Institute of Technology, Cambridge, MA 02139 USA (e-mail: jc\_woo@mit.edu; vipindev@mit.edu; medard@mit.edu).}
\thanks{Benjamin D. Kim is with the Department of Electrical and Computer Engineering, University of Illinois Urbana-Champaign, Champaign, IL 61820 USA (e-mail: bdkim4@illinois.edu).}
\thanks{Rafael G. L. D'Oliveira is with the School of Mathematical and Statistical Sciences, Clemson University, Clemson, SC 29631 USA (e-mail: rdolive@clemson.edu).}
\thanks{Alejandro Cohen is with the Technion Faculty of Electrical Engineering, Technion Israel Institute of Technology, Haifa 32000, Israel (e-mail: alecohen@technion.ac.il).}
\thanks{Thomas Stahlbuhk is with MIT Lincoln Laboratory, Lexington, MA 02421 USA (e-mail: thomas.stahlbuhk@ll.mit.edu).}
\thanks{Ken R. Duffy is with the Department of Electrical \& Computer Engineering and the Department of Mathematics, Boston University, Boston, MA 02215 USA (e-mail: k.duffy@northeastern.edu).}
}

\maketitle

\begin{abstract}
The Internet of Things (IoT) represents a significant advancement in digital technology, with its rapidly growing network of interconnected devices. This expansion, however, brings forth critical challenges in data security and reliability, especially under the threat of increasing cyber vulnerabilities. Addressing the security concerns, the Advanced Encryption Standard (AES) is commonly employed for secure encryption in IoT systems. Our study explores an innovative use of AES, by repurposing AES padding bits for error correction and thus introducing a dual-functional method that seamlessly integrates error-correcting capabilities into the standard encryption process. The integration of the state-of-the-art Guessing Random Additive Noise Decoder (GRAND) in the receiver's architecture facilitates the joint decoding and decryption process. This strategic approach not only preserves the existing structure of the transmitter but also significantly enhances communication reliability in noisy environments, achieving a notable over 3 dB gain in Block Error Rate (BLER). Remarkably, this enhanced performance comes with a minimal power overhead at the receiver—less than 15\% compared to the traditional decryption-only process, underscoring the efficiency of our hardware design for IoT applications. This paper discusses a comprehensive analysis of our approach, particularly in energy efficiency and system performance, presenting a novel and practical solution for reliable IoT communications.

\end{abstract}

\begin{IEEEkeywords}
Internet of Things (IoT), advanced encryption standard (AES), forward error correction (FEC), hardware architecture design, communication system reliability, energy efficiency in IoT\end{IEEEkeywords}

%
\IEEEpeerreviewmaketitle

\section{Introduction}

%
%
%
%
\IEEEPARstart{T}{he} Internet of Things (IoT) is a pivotal force in the evolution of digital technology, marked by the rapid integration of interconnected devices into a wide array of sectors~\cite{li2015internet}. These applications, ranging from smart homes to advanced traffic systems in transportation, remote health monitoring in healthcare, smart farming in agriculture, and supply chain optimization in business, are revolutionizing traditional practices~\cite{asghari2019internet}. They improve efficiency, precision, and decision-making capabilities, reflecting the profound impact of IoT on everyday life. As the IoT network continues to grow, with projections suggesting a leap to billions of connected devices, the challenges of data security and privacy have escalated. In this interconnected environment, where data is continuously exchanged, securing these vast streams of information becomes a critical task~\cite{7405513}.



In the context of IoT, the use of lightweight cryptographic techniques such as the Advanced Encryption Standard (AES)~\cite{daemen1999aes} is essential for securing data in energy-constrained environments. AES provides a practical balance, aligning with the dual demands of security and operational efficiency in IoT devices. It represents a strategic choice for IoT applications, ensuring that security measures are in harmony with the inherent constraints of these devices. AES, being a block cipher, always operates over a fixed block size of 128 bits (16 bytes). AES encryption typically incorporates padding to fill the remaining bits of its block in case of a mismatch in the input size~\cite{frankel2003aes}. Standards and protocols that are prevalent in IoT applications, such as IEEE 802.15.4~\cite{9144691} or Message Queuing Telemetry Transport (MQTT)~\cite{locke2010mq}, often employ such padding bits to match the AES block sizes. 

Recent advances in communication research have highlighted a unique aspect of AES, specifically its application in error correction~\cite{10149008}. This exploration revealed that the portion of an AES block used for padding can effectively provide error correction capabilities, thereby opening avenues for additional functionalities beyond its primary role in encryption. Building upon this insight, this work aims to explore the potential of the existing AES encryption and its padding scheme to improve communication reliability in IoT devices. By multi-purposing the existing padding bits required for AES encryption for error correction, we introduce a method that infuses error-correcting capabilities into the standard encryption process. This approach retains the original transmitter-side processes intact, while embedding an additional layer of error correction to improve data transmission reliability. 

In this research, the primary aim is to utilize AES's encryption capabilities and padding strategy to enhance communication reliability in IoT devices, without altering its existing security standards. Our method ingeniously repurposes existing AES padding bits for error correction, significantly enhancing data transmission reliability without necessitating modifications to the transmitter's architecture. This strategic application of AES,without necessitating operational alterations, fortifies communication against errors, evidenced by notable improvements in BER and BLER across diverse scenarios. This is complemented by a comparative analysis with the state-of-the-art Guessing Random Additive Noise Decoder (GRAND)~\cite{8630851} in practical application scenarios, assessing the method's effectiveness in realistic communication environments.

We have also implemented and compared the hardware for the state-of-the-art approaches as the baseline system and our proposed AES-based error correction architecture. This comparison focused on key metrics such as power consumption, latency, goodput, decoding energy required per bit (energy/bit), and hardware area. Through this comprehensive evaluation, we highlight the practical benefits of our approach, especially in terms of energy efficiency and system performance in real-world IoT applications.

This study aligns with scenarios outlined in the IEEE 802.15.4 standard, where Forward Error Correction (FEC) is usually not employed due to energy constraints, and retransmissions are used for error handling~\cite{9144691}. Our method of utilizing AES padding for error correction provides an efficient way to enhance system reliability and reduce retransmission rates in noisy channels. This offers a dual benefit: it provides FEC without necessitating additional changes to the transmitter side and provides better transmission rates as per IEEE standards.

The primary contributions of our study are summarized as follows:
\begin{itemize}
    \item Introduction of AES-based error correction in existing transmitter systems for IoT applications to enhance communication reliability in noisy environments, without the need for additional encoding processes.

    \item Comparative analysis of error correction capabilities between our AES-based method and the GRAND decoder, focusing on practical communication scenarios to demonstrate the effectiveness of our approach.

    \item Implementation and evaluation of hardware for both the baseline and AES-based error correction systems, analyzing power consumption, throughput, decoding energy required per bit, and hardware area.

    \item Demonstration of improvements in energy efficiency and overall system performance in IoT communication systems, emphasizing key performance indicators such as energy/bit and throughput.
\end{itemize}

The remainder of the paper is organized as follows: Section \ref{sec:previous_work} discusses related works on AES in error correction and its relevance to IoT systems. Section \ref{sec:proposed_system} details our proposed system, covering both the transmitter and receiver aspects. Section \ref{sec:hardware_design} details the hardware design for receiver systems in IoT devices, comparing our novel architecture with different traditional receiver architectures. System analysis, including error-correcting capabilities and hardware evaluation, is presented in Section \ref{sec:system_analysis}. Section \ref{sec:discussion} offers a discussion on the broader implications and potential future research directions. Finally, the paper concludes with Section \ref{sec:conclusion}, summarizing the key findings and their significance.

\section{Related Works} \label{sec:previous_work}

\subsection{The Internet of Things}
The phrase Internet of Things (IoT) was coined by Kevin Ashton \cite{ashton2009internet} in the late 1990's to represent the wide range of sensors and other small digital devices that capture and process data and gained a lot of interest in the early 2000s onwards. With the rapid advancements in technology, IoT devices that are digitally identifiable, with data capturing and processing capabilities, and connected to the internet have dominated the connected world. However, these resource-constrained devices are also prone to security threats and require lightweight cryptosystems to operate efficiently \cite{adat2018security,7562568}. There are different communication protocols defined for secure and reliable IoT operations such as the IEEE 802.15.4 and MQTT. The IoT devices, characterized by low data rates and their limited storage and processing capabilities, often prioritize encryption over channel coding schemes. For example, the IEEE 802.15.4 standard for ZigBee~\cite{alliance2008zigbee} proposes AES as the encryption scheme and depends on a Cyclic Redundancy Check (CRC), a scheme that checks for error detection, while relying on retransmissions of lost packets for reliable communication \cite{9144691}. This leads to an increased active period for the devices resulting in higher energy consumption, especially in lossy channels. Different performance analysis studies using the IEEE 802.15.4 standard focus on Eb/N0 up to 10 dB and achieve a Bit Error Rate (BER) of $10^{-4}$ in that range \cite{6749420,alnuaimi2006performance}. Even though not part of the standards, incorporating channel coding techniques such as Reed Solomon codes to IoT scenarios to reduce retransmissions and thus improve efficiency has been studied \cite{6663775,saadon2013evaluating}. However, these schemes require significant modifications to devices, including the integration of an encoder module, as well as the implementation of a corresponding decoding process at the receiver. This leads to increased power consumption and elevated computational complexity, presenting particular challenges for IoT devices where energy efficiency and processing power are paramount concerns.

\subsection{Advanced Encryption Standard}

The Advanced Encryption Standard (AES) is widely recognized as a robust symmetric block cipher scheme, mainly used to protect sensitive data~\cite{daemen1999aes}. Characterized by a fixed block size and supporting key sizes of 128, 192, and 256 bits, AES is known for its thorough encryption process. This process involves multiple rounds of encryption - 10, 12, or 14, depending on the key size - where each round, except the last, includes four distinct steps: SubBytes, ShiftRows, MixColumns, and AddRoundKey. These steps intricately mix the data with a key derived for each round, ensuring a high level of security.

A key aspect of AES, particularly relevant to our research, is its output's statistical randomness. AES has been effectively analyzed as a pseudo-random number generator, a characteristic that offers valuable potential beyond its primary encryption role~\cite{10.1145/945511.945515}. Due to this inherent randomness of AES output, it becomes viable to utilize AES for generating error-correcting performance akin to Random Linear Codes (RLCs)~\cite{10149008}. In leveraging this property, we explore the dual utility of AES - not only as an encryption mechanism but also as an effective means for error correction. This innovative application highlights the versatility of AES in enhancing communication reliability, particularly in scenarios where error correction is paramount.

\subsection{Guessing Random Additive Noise Decoding}

Guessing Random Additive Noise Decoding (GRAND) is a decoding algorithm employed at the receiver to interpret corrupted binary sequences~\cite{8437648, 8630851}. The GRAND process works by generating putative noise effect sequences based on channel conditions or soft data inputs. These sequences are ordered from the most likely to the least likely, using criteria such as Hamming weight for binary symmetric channels or logistic-weight for systems with soft input like per-bit reliabilities~\cite{6772729, 9872126}. This process is effectively illustrated in Algorithm \ref{algorithm:GRAND}, where each potential noise effect is tested against the received sequence for codebook membership. This methodology allows for maximum likelihood decoding and highlights the flexibility of GRAND as a universal decoder that works for any encoding method by verifying codebook membership.

ORBGRAND, an extension of GRAND, is optimized for soft detection scenarios and enhances block error rate (BLER) performance~\cite{9872126}. It processes reliability bits in a specific order, derived from soft information, for more effective decoding of block codes. This variant adapts the GRAND approach to accommodate soft input scenarios, such as those involving per-bit reliabilities, by generating noise effects based on increasing logistic-weight. ORBGRAND thus bridges the gap between the hard detection efficacy of GRAND and the complexity of soft detection techniques, proving essential for ultra-reliable, low-latency communications in environments like IoT.

\begin{algorithm}
    \caption{Guessing Random Additive Noise Decoding}
    \begin{algorithmic}[1]
        \State \textbf{Inputs:} A demodulated channel output \(y^n = (y_1, y_2, \ldots, y_n)\); a code-book membership function such that \(C(y^n) = 1\) if and only if \(y^n\) is in the code-book; and optional statistical noise characteristics or soft information, \(\Phi\).
        \State \textbf{Output:} decoded element \(c^n_*\).
        \Procedure{Decoding}{$C, y^n, \Phi$}
            \State \(d \gets 0\)
            \While{\(d = 0\)}
                \State \(z^n \gets\) next most likely binary noise effect sequence (which may depend on \(\Phi\))
                \If{\(C(y^n \oplus z^n) = 1\)}
                \State \(c^n_* \gets y^n \oplus z^n\)
                \State \(d \gets 1\)
                \EndIf
            \EndWhile
            \State \Return \(c^n_*\)
        \EndProcedure
    \end{algorithmic}
    \label{algorithm:GRAND}
\end{algorithm}
\vspace{-0.5 cm}

\section{Proposed System} \label{sec:proposed_system}

In this section, we delve into IoT system scenarios with a focus on communication reliability. We consider cases where encryption is prevalent, necessitating padding to align with AES's 128-bit block size. Our design maintains the encryption framework at the transmitter end while introducing an innovative hardware architecture for joint decoding and decryption at the receiver, utilizing GRAND as proposed in~\cite{Duffy_2019,10067519}. Our analysis contrasts our proposed system with traditional setups: a baseline system employing only encryption, thus lacking error correction, and another that separates encryption and encoding. The latter's receiver architecture includes distinct decryption and decoding phases. This comprehensive evaluation allows us to compare the error-correcting capabilities and decoding energy efficiency of our AES-based system with both the baseline encryption-only method and separated encryption-encoding systems. Detailed insights into each system's performance in various operational scenarios are provided in the following subsections.




\subsection{Transmitter}
In this section, we investigate three scenarios, each addressing how the payload is processed for reliable transmission.

\subsubsection{Padding for encryption, no redundancy bits (Baseline)} \label{tx_encrypt_only}
\begin{figure}
    \centering
    \includegraphics[width=1\linewidth]{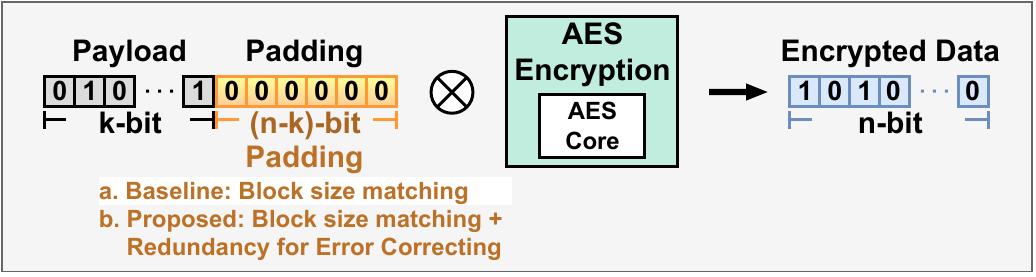}
    \caption{Transmitter: Encryption only. a. Baseline, b. Proposed}
    \label{fig:Tx_Encrypt_only}
\end{figure}
Figure \ref{fig:Tx_Encrypt_only} shows a standard encryption setup, adding padding to a $k$-bit payload to reach the $n$-bit AES block size, followed by encryption. However, these padding bits are only used for data security and do not provide error correction capabilities in traditional IoT systems.

\subsubsection{Padding for encryption, separate redundancy bits for error correction} \label{tx_encrypt_and_encode}
\begin{figure}
    \centering
    \includegraphics[width=1\linewidth]{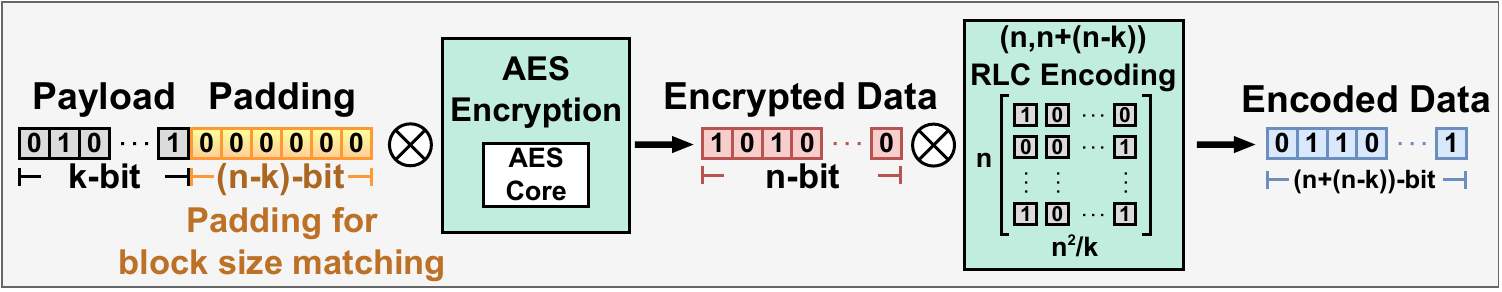}
    \caption{Transmitter: Separate encryption and encoding}
    \label{fig:Tx_Encrypt_and_Encode}
\end{figure}
In Figure \ref{fig:Tx_Encrypt_and_Encode}, the depicted process adds encoding after encryption, a typical approach in conventional secure and reliable communication systems~\cite{day1983osi, Bauer2013}. This step incorporates redundancy bits that enhance the error-correcting capability, thereby improving the reliability of the channel. Our comparison between this architecture and our proposed one will primarily focus on hardware aspects, such as power consumption, area, and energy efficiency. This allows us to demonstrate the practicality and effectiveness of our system in real-world applications.


\subsubsection{Padding for encryption, use the padding for error correction (Proposed)} \label{tx_proposed}
In the proposed system, as outlined in Figure~\ref{fig:Tx_Encrypt_only}, the transmitter-side architecture remains unchanged from the first scenario. The $k$-bit payload undergoes padding with $(n-k)$ bits to align with the $n$-bit AES block size. However, what distinguishes this approach is the alteration in the receiver architecture, where these padding bits are now utilized for error correction. This is achieved through the novel integration of the GRAND joint decoding and decryption functionality, offering an innovative method for enhancing error correction without modifying the transmitter's setup.

\subsection{Receiver}
Our research employed both performance simulations and hardware implementations to validate our approach, ensuring its applicability in energy-constrained IoT environments. The study involved a detailed comparison of three receiver architectures, each designed to align with specific transmitter setups. These architectures, focusing on reliable and efficient decoding in challenging channel conditions, were critical for understanding their viability in IoT systems, where energy efficiency and effective data processing are paramount.

\subsubsection{AES decryption without decoding (Baseline)} 
The baseline receiver system, corresponding to the ``Padding for encryption, no redundancy bits" transmitter setup outlined in Section \ref{tx_encrypt_only}, is designed solely for AES decryption, without any error-correction capability. This baseline configuration is crucial for our study, providing a standard against which we can evaluate the error-correcting efficiency of our proposed system. Our analysis not only compares the reliability and efficiency improvements of our system over this baseline but also includes an assessment of the hardware overheads involved in implementing our advanced error-correcting approach in the receiver. This comparison is key to highlighting the practical benefits and potential trade-offs of our system in IoT environments.

\subsubsection{Separate ORBGRAND decoding and AES decryption}
\label{rx_separate}
\begin{figure*}
    \centering
    \includegraphics[width=1\linewidth]{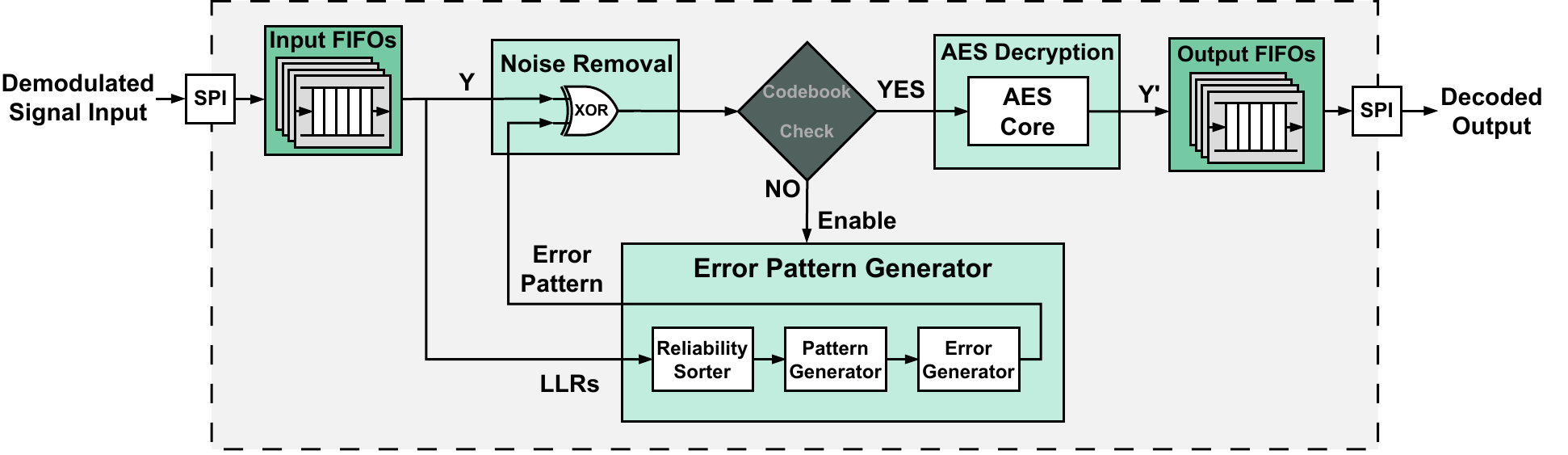}
    \caption{Receiver: separate ORBGRAND decoding and AES decryption}
    \label{fig:rx_separate_FIFO}
\end{figure*}

In the receiver architecture aligned with Section \ref{tx_encrypt_and_encode}, titled ``Padding for encryption, Redundancy bits for error correction," the approach involves separate ORBGRAND decoding and AES decryption. As depicted in Figure~\ref{fig:rx_separate_FIFO}, this process starts with an attempt at codebook matching. If the match fails, the system activates the error pattern generator, which operates based on the reliability of received bits. An error pattern is generated, subtracted from the received signal, and codebook matching is retried until a match is found. Following successful decoding, the AES decryption module decrypts the decoded block.

\subsubsection{ORBGRAND + AES joint decoding and decryption (Proposed)} \label{rx_proposed}

\begin{figure*}
    \centering
    \includegraphics[width=1\linewidth]{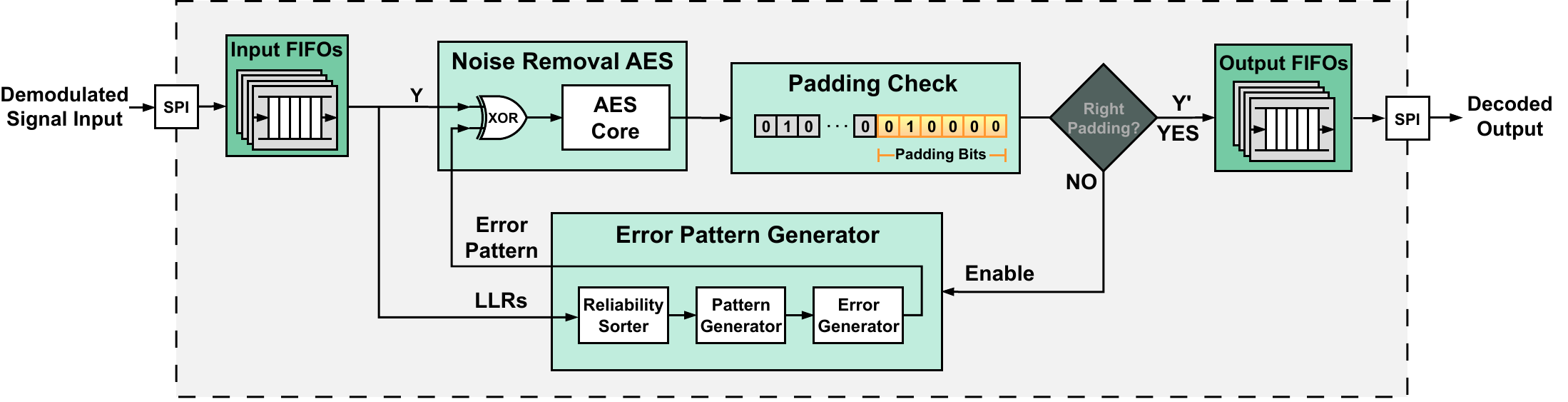}
    \caption{Receiver: ORBGRAND + AES joint decoding and decryption}
    \label{fig:Rx_Proposed_FIFO}
    \vspace{-0.5 cm}
\end{figure*}

The proposed architecture in Figure \ref{fig:Rx_Proposed_FIFO} is tailored to~\ref{tx_proposed} ``Padding for encryption, use padding for error correction" transmitter scenario. It begins with AES decryption, checking the padding sequence. Upon detecting padding sequence discrepancies, the system activates the error pattern generator. It iteratively adjusts the received signal based on generated error patterns until the correct padding sequence is detected post-decryption.

In comparing these architectures, the proposed system, with its integrated decoding and decryption, aims to reduce latency, whereas the separate decoding and decryption system may involve more clock cycles due to its sequential approach. The proposed system also occupies a smaller area, as it omits the need for a codebook checking matrix. These aspects, particularly in terms of latency, energy demands, and area efficiency will be comprehensively analyzed in Section~\ref{sec:system_analysis}, taking into account the constraints of energy-efficient IoT systems.

\section{Energy-Efficient Hardware Design} \label{sec:hardware_design}
The design and implementation of the receiver systems in IoT devices is a meticulous process, especially when focusing on the baseline receiver system, which comprises the ORBGRAND decoder followed by AES decryption, and our novel architecture integrating AES for error correction with joint decoding and decryption.

\subsection{Error Pattern Generator}
Our receiver hardware's error pattern generator design strategically utilizes a 128-bit block size. This choice, deviating from the larger block sizes found in previous designs~\cite{10067519}, enhances hardware efficiency. The smaller block size reduces the number of comparator stages needed, thereby lowering the clock cycles required for generating error patterns, speeding up the process and boosting overall efficiency. The optimized error pattern generator's adoption is consistent across both the separate decoding and decryption receiver architecture and our proposed receiver architectures, ensuring uniform performance across different system configurations.

\subsection{Decoding Process in Proposed Receiver Hardware}
The decoding process in our proposed receiver hardware architecture is strategically designed to improve system performance:

\begin{enumerate}
    \item \textbf{Data Fetch and Storage:} This initial stage handles the retrieval and storage of the incoming data, setting the foundation for the decoding process.
    \item \textbf{Decryption and Padding Check:} Combining decryption with padding check, this stage employs AES-128\footnote{This study employs AES-128 for its reduced operational rounds compared to AES with larger key sizes. Nevertheless, the approach can be adapted for different AES key sizes.} for its efficiency. Completing in just 13 clock cycles, it includes data input and output, ensuring a continuous flow of data.
    \item \textbf{Error Pattern Generation and Noise Removal:} If discrepancies in the padding sequence are detected, the system generates and subtracts error patterns based on the reliability of the received bits, correcting potential transmission errors.
    \item \textbf{Iterative Decryption and Padding Check:} The process of decryption and padding check repeats until the correct padding sequence is identified, ensuring the accuracy and reliability of the decoded data.
\end{enumerate}

Incorporating AES-128, selected for its compact area and efficient energy usage per bit, is a critical component of our design~\cite{8721457}. This pipelined architecture is especially advantageous in high SNR conditions, effectively reducing latency to the duration of a single decryption cycle. By focusing on minimizing clock cycles for error pattern generation and leveraging a pipelined architecture, we have created a balance between the need for error correction and the demand for energy efficiency, a key consideration in the realm of IoT systems constrained by resource limitations.

\section{System Analysis} \label{sec:system_analysis}

This section presents the results of evaluating the performance of AES as an error-correcting code in comparison to different transmitter setups discussed in Section~\ref{sec:proposed_system}. We also include hardware simulation results for the receiver architecture, focusing on power, throughput, energy per bit, and area. The receiver system was designed and implemented using 28nm CMOS technology, and the results were obtained from post-synthesis simulations at 0.9V and 100MHz.

\subsection{Error Correcting Capability}
In our error-correcting performance evaluation, binary messages of length k-bits, chosen uniformly from $F^{k}_{2}$, were transmitted using Binary Phase Shift Keying (BPSK) over an Additive White Gaussian Noise (AWGN) channel. We maintained an equal number of redundancy bits across different transmitter encoding scenarios. This approach was essential since the error-correcting performance of the GRAND decoder is influenced by the number of redundancy bits~\cite{Duffy_2019}. Performance metrics such BER and BLER were analyzed based on the energy per information bit to noise power spectral density ratio ($Eb/N_{0}$)~\cite{enwiki:1121091972}.

In our simulations, we investigated three distinct scenarios:
\begin{itemize}
    \item \textbf{Baseline Encryption-Only without Encoding:} This system focuses solely on encryption, applying hard decoding where any bit error leads to decoding failure. It is a straightforward approach emphasizing data security, without incorporating any additional encoding.
    \item \textbf{Separate Encryption and Encoding:} This setup uses ORBGRAND for decoding, leveraging soft information from signal reliability. It combines encryption with subsequent encoding, enhancing error correction while maintaining secure data transmission. The choice of ORBGRAND is due to its superior error-correcting capabilities, flexibility, and energy efficiency in hardware, making it a suitable option for this architecture~\cite{10067519}.
    \item \textbf{Proposed AES as Error Correction:} This method innovatively combines joint decoding and decryption using ORBGRAND. It enhances error correction capabilities within the framework of AES encryption, providing a balanced approach to secure and reliable data transmission.
\end{itemize}

\begin{figure*}
    \centering
    \includegraphics[width=1\linewidth]{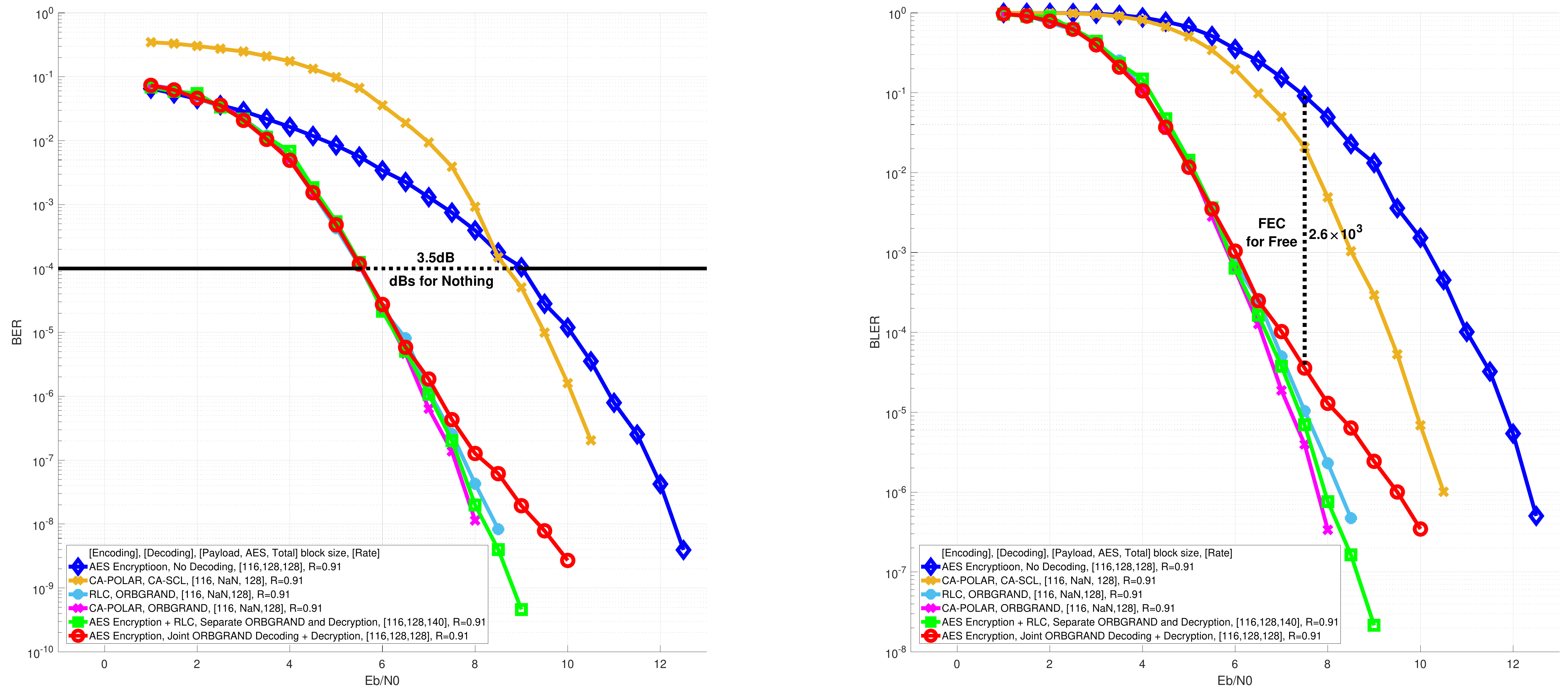}
    \caption{BER (left) and BLER (right) vs. $Eb/N_{0}$ for padding bits = 12-bit}
    \label{fig:(128,116)_results}
\end{figure*}

\begin{figure*}
    \centering
    \includegraphics[width=1\linewidth]{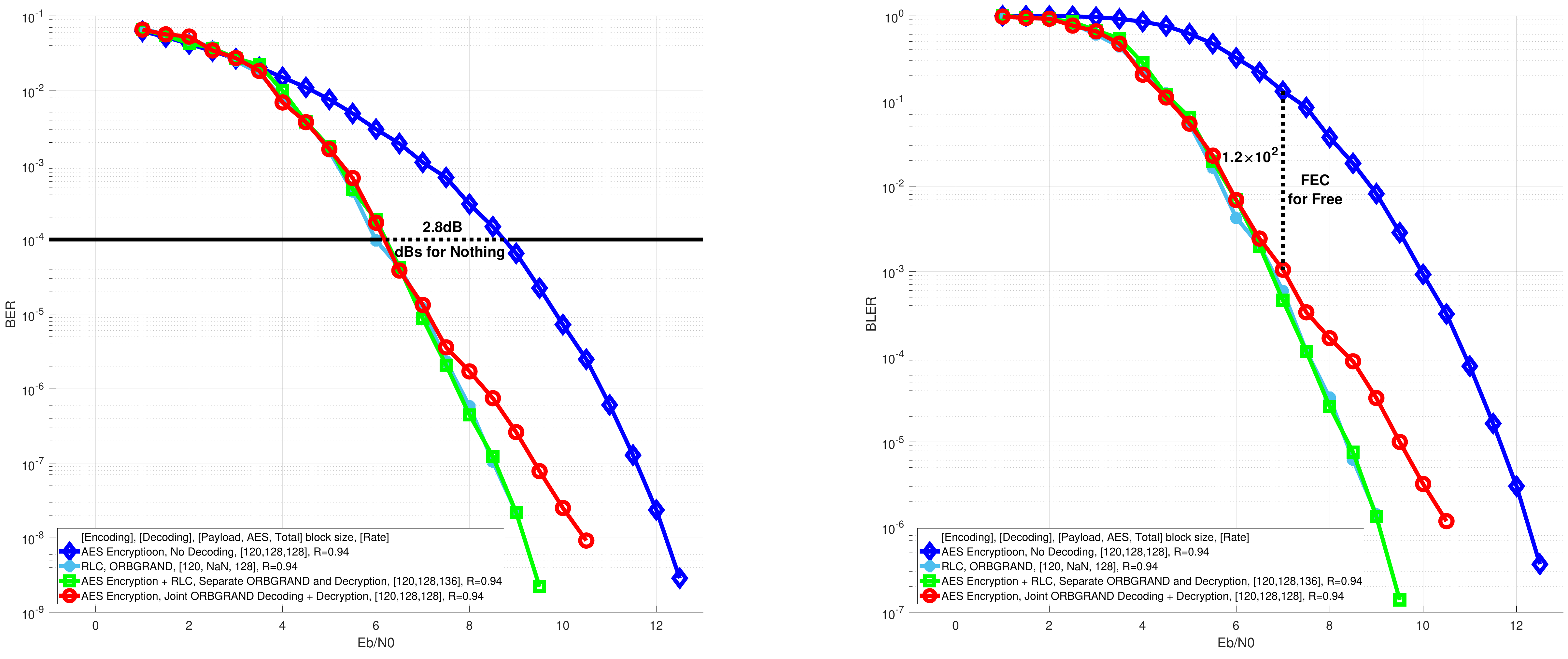}
    \caption{BER (left) and BLER (right) vs. $Eb/N_{0}$ for padding bits = 8-bit}
    \label{fig:(128,120)_results}
    \vspace{-0.5 cm}
\end{figure*}


Figures~\ref{fig:(128,116)_results} and~\ref{fig:(128,120)_results} provide a comprehensive analysis of BER and BLER against $Eb/N{0}$. These figures showcase different scenarios with 8-bit and 12-bit padding. The proposed AES error correction scheme significantly improves over the baseline encryption-only system, especially with 12-bit padding. This improvement is particularly remarkable in the BER under $10^{-4}$, indicating a substantial improvement in the reliability of the system.

The first set of trials with 12-bit padding demonstrated the robustness of our AES error correction strategy. It was compared with state-of-the-art encoding and decoding schemes such as RLC~\cite{article} and CA-Polar~\cite{7055304} for encoding, and ORBGRAND~\cite{9872126} and CRC-Aided Successive Cancellation List (CA-SCL)~\cite{6297420} for decoding. Our approach not only outperformed the ``no encoding" scenario but also showed competitive capabilities against the widely used "CA-POLAR" code with CA-SCL decoding. Significantly, this level of error correction effectiveness was achieved with less than 10\% of the AES block size as padding bits, underscoring the efficiency of our method. Additionally, it is noteworthy that ORBGRAND displayed significantly superior performance over the state-of-the-art CA-SCL decoder in both BER and BLER metrics, further establishing its effectiveness.

In scenarios involving 8-bit and 12-bit padding, our proposed system consistently outperformed the baseline encryption-only setup, demonstrating its versatility across different conditions. When compared with systems employing separate encryption and encoding — where encryption is followed by encoding — our integrated approach maintained similar levels of BER and BLER, if the number of redundancy bits is matched to the padding. This performance comparability, despite different operational methodologies, highlights the effectiveness of our proposed method in enhancing communication reliability without altering the security mechanism. Our approach maintains the same security standards provided by the AES with inherent padding while significantly improving reliability, demonstrating a balanced advancement in secure and reliable IoT communications.



In high $Eb/N_{0}$ scenarios, our AES-based error correction system experiences a marginal decrease in performance when compared to the separate encrypt-and-encode system, primarily due to AES's non-linear codebook characteristics. Unlike linear codes with uniform minimum Hamming distances~\cite{wei1991generalized}, AES's codebook does not maintain this uniformity, particularly when single-bit errors are more likely at these higher $Eb/N_{0}$ levels. In the AES system, due to its non-linear codebook, a single-bit error for certain codewords may result in an erroneous codeword being closer, leading to potential decreases in error correction efficacy. Simultaneously, our study leverages ORBGRAND, known for its near-optimal performance in moderate $Eb/N_{0}$ regimes, especially below 9 dB~\cite{9992258}. This is our ideal operation regime since no additional coding is required to maintain a BER below $10^{-4}$ for $Eb/N_{0}$ above 9 dB. The differences in the construction of AES and RLC, and the specific characteristics of ORBGRAND, are key factors in our system's performance, particularly under varying $Eb/N_{0}$ conditions. This combined understanding of ORBGRAND's capabilities and AES's inherent codebook properties clarifies the observed performance trends in different $Eb/N_{0}$ environments. This research emphasizes the effective application of ORBGRAND within the moderate $Eb/N_{0}$ regimes, without delving into its alternative operational modes that might be more suitable in higher $Eb/N_{0}$ scenarios since that is outside our area of interest.

Despite this minor decline in high $Eb/N_{0}$ conditions, our proposed system consistently outperformed the encryption-only approach in terms of lower BER and BLER. Achieving a BER better than $10^{-4}$ is particularly significant in practical use cases of IEEE 802.15.4 standard~\cite{6749420,alnuaimi2006performance}, underscoring the practicality and applicability of our system in real-world settings. This finding highlights the potential of our proposed AES error correction scheme in enhancing the reliability of communication systems, particularly in IoT environments where efficient data transmission is crucial.


\subsection{Hardware Analysis}
\begin{table*}[]
\resizebox{\textwidth}{!}{%
\renewcommand{\arraystretch}{2}
\begin{threeparttable}
\centering    
\caption{\centering Performance comparison in different receiver architectures \\ (Goodput is valid for BER $< 10^{-4}$)}
\label{tab:hardware_results}
\begin{tabular}{c}
    \includegraphics{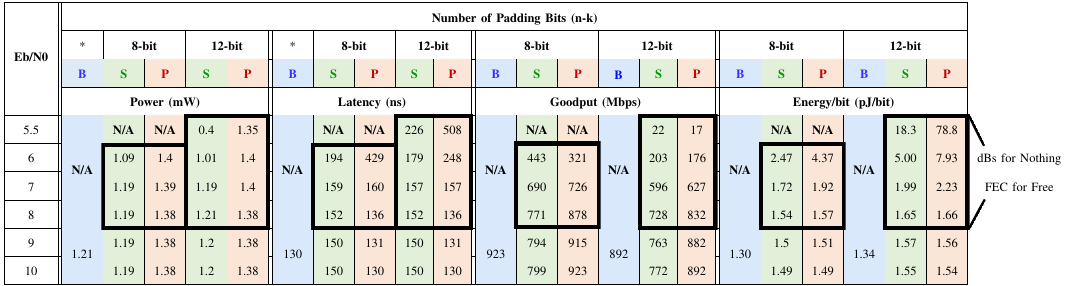}
\end{tabular}
     \begin{tablenotes} \footnotesize
       \item {\color[HTML]{3531FF}B: Baseline receiver system, AES decryption only}
       \item {\color[HTML]{009901}S: Separate receiver system, ORBGRAND decoding followed by AES decryption}
       \item {\color[HTML]{CB0000}P: Proposed receiver system, joint ORBGRAND decoding and AES decryption}
       \item {\textbf{N/A}: Unavailable due to high BER in the initial transmission}
     \end{tablenotes}
  \end{threeparttable}}
\end{table*}

\begin{figure}
    \centering
    \includegraphics[width=1\linewidth]{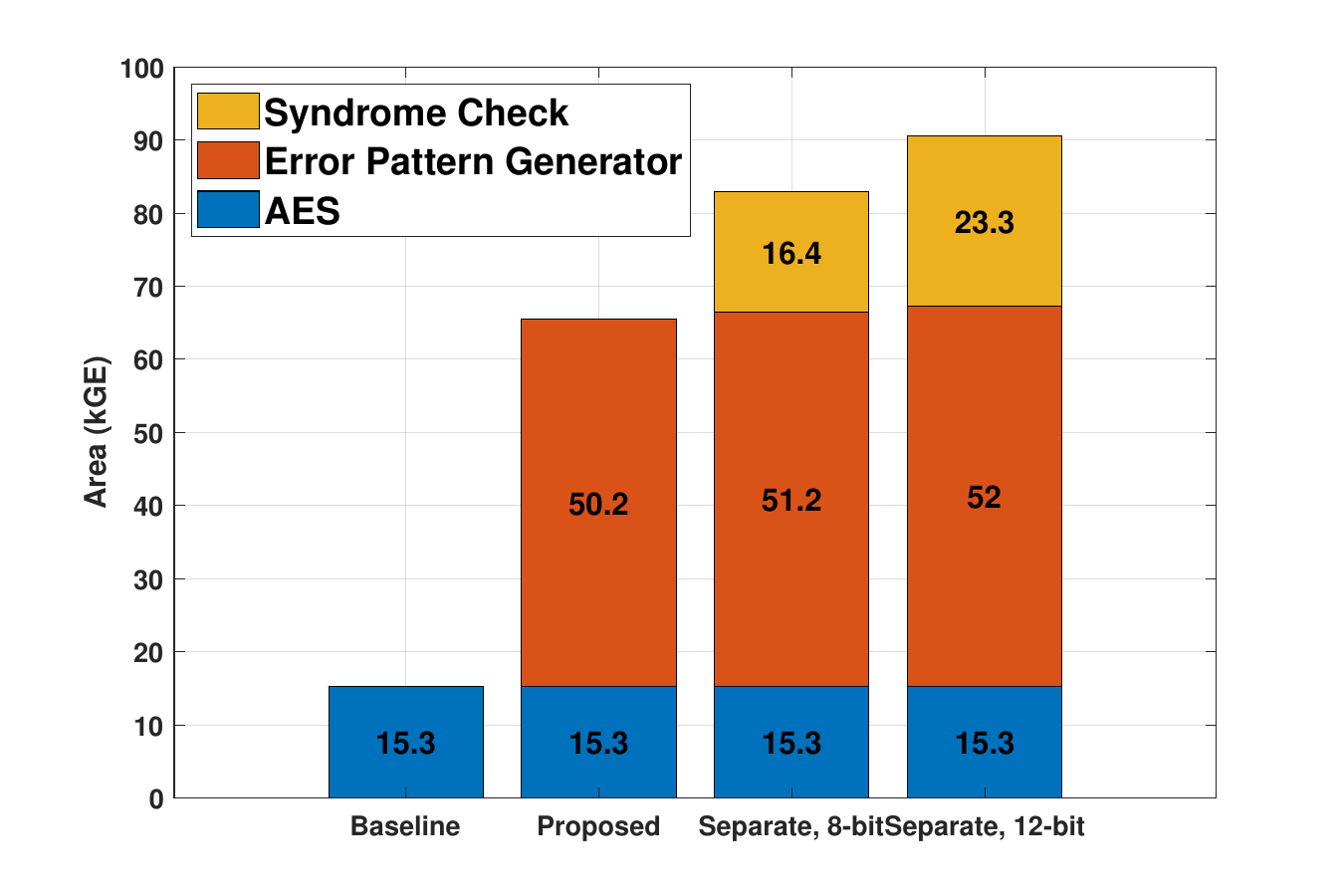}
    \caption{Area comparison in different receiver architectures. Proposed(left), separate with 8-bit redundancy(middle) and separate with 12-bit redundancy(right)}
    \label{fig:area}
\end{figure}

The hardware analysis in our study compares the performance of three different receiver architectures: our proposed system with integrated joint ORBGRAND and AES decoding and decryption, a baseline system focusing only on AES decryption without error correction, and a separate system that combines ORBGRAND decoding with subsequent AES decryption. This comprehensive analysis, detailed in Table~\ref{tab:hardware_results}, examines power consumption, latency, goodput, and energy efficiency under various $Eb/N_{0}$ scenarios along with different numbers of padding bits. The findings provide valuable insight into the operational efficiencies and energy demands of these systems, providing a clear understanding of their relative performances in different signal-to-noise ratio environments.

It's essential to note the baseline system's uniform operation across all noise levels, demonstrating consistent power and latency due to its sole reliance on AES decryption. However, the evaluation of goodput and energy per bit becomes particularly relevant under conditions where the baseline system's BER is maintained below $10^{-4}$, typically at $Eb/N_{0}$ values greater than 9. Instances not meeting this criterion, such as the $Eb/N_{0}$ = 5.5 dB scenario with 8-bit padding, are thus marked as N/A, underscoring the importance of maintaining a threshold for reliable transmission.

The baseline system, solely relying on AES decryption, displays consistent power consumption and latency across all noise levels, maintaining uniform operation regardless of channel conditions. However, its goodput and energy per bit, tied to the actual payload data ($k$-bit), are influenced by the size of the padding bits, thus varying with the number of redundancy bits used. This variation in goodput and energy efficiency reflects the baseline's dependence on payload size in different operational scenarios. Moreover, the system's goodput and energy efficiency metrics are directly linked to the actual payload data ($k$-bit) and are affected by the padding bits' size, indicating a dependency on payload size for different operational scenarios.

Our proposed receiver system experiences an additional power overhead from the error pattern generator. Remarkably, in high $Eb/N_{0}$ scenarios, this overhead is less than 15\% compared to the baseline decryption-only system, a rate that remains fairly consistent even in lower $Eb/N_{0}$ scenarios. The separate system, which employs sequential ORBGRAND decoding and AES decryption, demonstrates lower power consumption in lower $Eb/N_{0}$ scenarios. This is attributed to the system's operational strategy, where it runs the error pattern generator and syndrome check until successful decoding, followed by a single instance of AES decryption.

In the context of latency, our proposed system demonstrates no degradation compared to the baseline system in high $Eb/N_{0}$ scenarios, maintaining a consistent latency of 130ns for the entire decoding and decryption process, which is a direct result of the pipelined and parallel processing design. This performance is notably superior to the separate decoding and decryption system, which requires 150ns for the same process. The enhanced latency of our system translates into higher goodput, making it highly advantageous for IoT receiver applications where efficiency is key. However, in lower $Eb/N_{0}$ conditions, the need for frequent generation of new error patterns and padding checks in our system results in increased latency, which adversely impacts the goodput. Despite increased latency in lower $Eb/N_{0}$ conditions due to frequent new error pattern generation and padding checks, our system still manages to achieve a reliable communication channel with a BER under $10^{-4}$, balancing between latency and error correction efficiency in varying channel conditions.

In the high $Eb/N_{0}$ scenarios, the energy per bit metric for our proposed system demonstrates an overhead of less than 15\% compared to the baseline system. Despite the higher power demand of the proposed system, its superior goodput efficiency effectively compensates for this, resulting in a comparably efficient energy per bit performance. Furthermore, when we compare the proposed system with the separate system, the energy per bit figures are comparable across the two architectures, particularly notable in high $Eb/N_{0}$ conditions. In contrast, in low $Eb/N_{0}$ scenarios, despite having similar power consumption levels, the proposed system's reduced goodput leads to a relative increase in energy per bit, highlighting the impact of operational efficiency on energy metrics in different noise environments.

The area comparison between our proposed receiver architecture and the separate systems is illustrated in Figure~\ref{fig:area}. In our design, both the AES and error pattern generator components are optimized for low latency, tailored to a 128-bit block size. This approach ensures a consistent AES area across different systems. The area of the error pattern generator varies with the codeword length ($n$), but its basic architecture is designed to minimize latency in error pattern generation for all systems. Consequently, our proposed receiver architecture, while maintaining a similar AES area, benefits from a more compact error pattern generator due to the shorter codeword length, enhancing efficiency in space-constrained IoT applications.

The area efficiency of our proposed receiver system is enhanced due to the absence of a large $H$-matrix for syndrome check, a feature required in the separate systems. In the separate decoding and decryption system, the area for syndrome check is directly proportional to the product of the number of redundancy bits ($n-k$) and the codeword length ($n$), necessitating substantial storage in registers. Our system's approach, which involves padding check instead of syndrome check, eliminates the need for this extensive matrix storage. This design choice not only reduces the area by approximately 28\% compared to the separate systems but also becomes more beneficial at lower code rates. The absence of additional area requirements for padding check in our architecture emphasizes its suitability for IoT applications, where space and resource constraints are critical factors.

The overhead of area in our proposed system is mainly due to the error pattern generator, which is 50.2kGE (kilo-Gate Equivalent). However, this is still smaller than the separate system, which requires over 67kGE for error-correcting capabilities compared to a baseline receiver. This highlights the balance our system strikes between maintaining error-correcting capabilities and efficient use of space, vital in energy and area-constrained IoT environments.

\section{Discussion} \label{sec:discussion}
The findings of our study highlight the effective utilization of existing AES encryption in the transmitter setup for enhancing communication reliability, without necessitating changes to the transmitter's operation. By using existing AES padding bits for error correction, we significantly improve the reliability of the communication channel with minimal overhead in the receiver.

Our implementation hinges on the inclusion of an ORBGRAND error pattern generator in the receiver architecture. This addition incurs less than a 15\% power overhead but significantly improves the reliability of the communication channel. For instance, in the scenario depicted in Figure~\ref{fig:(128,116)_results}, aiming for a BER of $10^{-4}$, our proposed system achieved this BER at an $Eb/N_{0}$ of only 5.5 dB, 3.5 dB less than the 9 dB required by the no encoding scenario. Similarly, as shown in Figure~\ref{fig:(128,120)_results}, with channel $Eb/N{0}$ fixed at 7 dB, our system improved the BLER from $1.3\times10^{-1}$ in the no encoding scenario to a remarkable $1.0\times10^{-3}$. This efficiency not only validates the effectiveness of our approach but also indicates a substantial reduction in the required energy for the transmitter to achieve specific BER/BLER targets.

Another key advantage is the reduction in retransmission rates, particularly relevant to the IEEE 802.15.4 standard, which relies on retransmissions rather than FEC and adopts CRC for this purpose. In the context of Figure~\ref{fig:(128,116)_results} at $Eb/N{0}$ = 7.5 dB, our AES FEC approach can reduce retransmission rates from $9 \times 10^{-2}$ to as low as $3.5\times10^{-5}$, marking a significant improvement over the standard IEEE 802.15.4 communication error checking with CRC. Moreover, to achieve a similar error-correcting performance as our proposed AES FEC scheme, additional encoding after encryption would be required in the transmitter. This leads to increased bit transmission ($n$ vs. $n+(n-k)$), reducing throughput and requiring more transmission energy, highlighting the efficiency of our approach.


The flexibility of our method, including the potential for error checking and retransmission requests similar to CRC operations, adds robustness and adaptability to different operational needs and scenarios. This study demonstrates the feasibility and benefits of using AES for error correction in communication systems, particularly in IoT environments where energy efficiency and reliability are paramount.

\section{Conclusion} \label{sec:conclusion}
In conclusion, this research presents a novel approach to enhancing communication reliability in IoT systems by leveraging the existing AES encryption mechanism. Our proposed method repurposes the existing padding bits in AES encryption for error correction, effectively improving the reliability of data transmission without necessitating any changes to the transmitter setup. This innovative use of AES leaves the cryptographic consideration untouched while adding error-correcting functionality, showing significant improvements in BER and BLER across various scenarios without compromising the inherent security provided by AES encryption.


Our comprehensive system analysis includes a comparison with the state-of-the-art ORBGRAND decoder and an in-depth evaluation of receiver architectures in terms of power consumption, latency, goodput, energy, and area efficiency. The proposed system's design ensures minimal power overhead and matches the latency of a decryption-only receiver in high signal-to-noise ratio environments, making it highly suitable for energy and area-constrained IoT applications.

Moreover, the study underscores the potential of this approach in reducing retransmission rates, aligning with the IEEE 802.15.4 standard, which typically relies on retransmissions for error handling. By providing an effective error correction mechanism, our system significantly reduces the need for retransmissions, enhancing overall communication efficiency.

In essence, this work demonstrates the feasibility and advantages of using inherent AES padding for error correction in IoT communication systems, highlighting its dual benefits in improving system reliability and reducing retransmission rates, thereby contributing to the advancement of efficient IoT technologies.




%





\ifCLASSOPTIONcaptionsoff
  \newpage
\fi



%



\bibliographystyle{IEEEtran}
\bibliography{references.bib}

%








\end{document}